\documentclass[sigconf]{acmart}
\usepackage{threeparttable}
\usepackage{csquotes}
\usepackage{multicol}
\usepackage{booktabs}
\usepackage{threeparttable}
\usepackage{graphicx}
\usepackage{enumitem}

\AtBeginDocument{%
  }

\copyrightyear{2025}
\acmYear{2025}
\setcopyright{acmlicensed}\acmConference[UbiComp Companion
'25]{Companion of the 2025 ACM International Joint Conference on Pervasive
and Ubiquitous Computing}{October 12--16, 2025}{Espoo, Finland}
\acmBooktitle{Companion of the 2025 ACM International Joint Conference on
Pervasive and Ubiquitous Computing (UbiComp Companion '25), October
12--16, 2025, Espoo, Finland}
\acmDOI{10.1145/3714394.3756239}
\acmISBN{979-8-4007-1477-1/2025/10}

\pdfoutput=1

\begin{document}

\title{Design and Challenges of Mental Health Assessment Tools Based on Natural Language Interaction}

\author{Yixue Cai}
\orcid{0009-0008-4913-4444}
\affiliation{%
  \institution{Nankai University}
  \department{College of Software}
  \city{Tianjin}
  \country{China}
}
\email{2213221@mail.nankai.edu.cn}

\author{Xiyan Su}
\orcid{0009-0009-4196-8150}
\affiliation{%
  \institution{Nankai University}
  \department{College of Software}
  \city{Tianjin}
  \country{China}
}
\email{2313131@mail.nankai.edu.cn}

\author{Dongpeng Yao}
\orcid{0000-0002-8201-825X}
\affiliation{%
  \institution{Nankai University}
  \city{Tianjin}
  \country{China}
}
\email{13651362566@163.com}

\author{Rongduo Han}
\orcid{0009-0007-5157-4660}
\affiliation{%
  \institution{Nankai University}
  \department{College of Software}
  \city{Tianjin}
  \country{China}
}
\email{1120240404@mail.nankai.edu.cn}

\author{Nan Gao}
\orcid{0000-0002-9694-2689}
\authornotemark[1]
\affiliation{%
  \institution{Nankai University}
  \department{College of Software}
  \city{Tianjin}
  \country{China}
}
\email{nan.gao@nankai.edu.cn}

\author{Haining Zhang}
\orcid{0009-0007-1320-3981}
\authornote{Corresponding authors}
\affiliation{%
  \institution{Nankai University}
  \department{College of Software}
  \city{Tianjin}
  \country{China}
}
\email{zhanghaining@nankai.edu.cn}









\renewcommand{\shortauthors}{Yixue Cai et al.}
\begin{abstract}

Mental health assessments are of central importance to individuals' well-being. Conventional assessment methodologies predominantly depend on clinical interviews and standardised self-report questionnaires. Nevertheless, the efficacy of these methodologies is frequently impeded by factors such as subjectivity, recall bias, and accessibility issues. Furthermore, concerns regarding bias and privacy may result in misreporting in data collected through self-reporting in mental health research. The present study examined the design opportunities and challenges inherent in the development of a mental health assessment tool based on natural language interaction with large language models (LLMs). An interactive prototype system was developed using conversational AI for non-invasive mental health assessment, and was evaluated through semi-structured interviews with 11 mental health professionals (six counsellors and five psychiatrists). The analysis identified key design considerations for future development, highlighting how AI-driven adaptive questioning could potentially enhance the reliability of self-reported data while identifying critical challenges, including privacy protection, algorithmic bias, and cross-cultural applicability. This study provides an empirical foundation for mental health technology innovation by demonstrating the potential and limitations of natural language interaction in mental health assessment.
\end{abstract}


  



\maketitle

\vspace{-1mm}

\section{Introduction}




Mental health assessments are imperative for the well-being and social functionality of individuals\cite{world2013comprehensive}. Conventional mental health assessment methodologies are mainly based on clinical interviews, professional counseling, and standardised self-report questionnaires. Despite the prevalence of these methods, they are frequently constrained by subjectivity, recall bias, and accessibility barriers. Nevertheless, a substantial body of research has cast doubt on the reliability of self-reporting. For instance, researchers have observed a semantic discrepancy between self-reported and observed emotions\cite{boyd2023automated}, as well as low consistency between self-reporting and behavioral measures. Longitudinal studies have further revealed issues such as social desirability bias, recall errors, and declining compliance over time \cite{adler2021identifying, chowdhury2025predicting, li2020extraction, shin2024using, xu2021leveraging}. These challenges are especially salient in the domain of mental health, as social stigmatization and concerns regarding privacy may result in intentional or unintentional misreporting.


Natural language interaction is a promising approach to address these limitations. It can reduce the artificiality of structured questionnaires while capturing richer, contextual information about an individual's mental state. Large language models (LLMs) are suitable for mental health applications where nuanced communication is essential.

This study explores the design of mental health assessment tools based on natural language interaction with LLMs. We have developed an interactive prototype system, the Intelligent Mental Health Tool, which uses conversational AI to conduct non-invasive mental health assessments. The system can generate and adapt self-report questions based on user feedback, analyse text data to infer emotional states and detect inconsistencies, and enhance psychological log analysis methods. Furthermore, it has been demonstrated that the platform provides users and clinicians with interpretable feedback, thus enhancing transparency and trust.

To evaluate and refine this approach, we conducted in-depth interviews with six professional counsellors and five psychiatrists. These interviews helped us map typical workflows in counselling sessions, determine the current state of AI integration in mental health services, and gather expert perspectives on the challenges and opportunities associated with LLM-based assessment tools. Our research was guided by the following questions:





\begin{quote}
How to design a mental health assessment tool based on natural language interaction? What are the main challenges it currently faces?
\end{quote}

This study offers an empirical foundation for mental health technology innovation and explores the potential of AI-driven adaptive questioning to improve the reliability of self-reported data. The study also highlights key challenges such as privacy protection, bias, and cross-cultural applicability.

The contributions of this paper are as follows.
\begin{itemize}
    \item Through in-depth interviews with professional counsellors and psychiatrists, we systematically explored the limitations and challenges of self-reported data in mental health assessment and identified specific requirements for LLM-based assessment tools;
    \item We proposed and evaluated a design methodology for an adaptive questionnaire system based on large language models, which can dynamically adjust questions, integrate multimodal data, and provide interpretable feedback;
    \item We identified critical challenges in implementing such systems, including privacy concerns, bias mitigation, and cross-cultural adaptation, providing concrete guidance for future research and development in this emerging field.
\end{itemize}

\section{Related Works}

\subsection{Measurements of Mental Health}

Traditional mental assessment methods are often limited by subjectivity, recall bias, and accessibility barriers \cite{stone2003patient}. However, recent technological advances have enabled new approaches, such as mobile devices and wearable sensors, to passively and continuously collect behavioural and physiological data in real-world settings \cite{cornet2018systematic} For example, Samyoun et al. \cite{samyoun2022m3sense} proposed M3Sense, a framework that uses multimodal sensors to capture user behaviour and impact, while Tlachac et al. \cite{tlachac2022deprest} demonstrated the use of smartphone logs for mental health screening during the pandemic. In educational contexts, Gao et al. (2020) predicted student engagement using the n-Gage system. These new methods, which combine machine learning and digital data, offer more granular and personalised models of mental health \cite{adler2021identifying, chowdhury2025predicting, li2020extraction, shin2024using, xu2021leveraging}, but also raise new challenges around privacy and data interpretation.

\subsection{The Challenge of Self-Report Authenticity}

A substantial body of research has cast doubt on the reliability of self-report. For instance, Boyd \& Andalibi \cite{boyd2023automated} observed a semantic gap between observed emotions and self-reported affect, while Das Swain et al. \cite{das2022semantic} found low agreement between self-report and behavioural measures. Longitudinal investigations have further revealed issues such as social desirability bias, recall errors, and declining compliance over time \cite{gao2020n, gao2021investigating, wash2017can}. These challenges are especially evident in mental health settings, where stigma and concerns about privacy can lead to deliberate and inadvertent misreporting.

\subsection{AI and LLM-based Solutions}

For example, Xu et al. \cite{xu2024mental} conducted a comprehensive evaluation of large language models on various mental health prediction tasks using online text data, highlighting the effectiveness of instructional fine-tuning in improving model performance. Bouguettaya et al. \cite{bouguettaya2025ai} provided a comprehensive review of AI-driven report-generation tools in mental healthcare. Furthermore, several studies have examined the effectiveness of LLMs as psychological assessors, demonstrating their potential to provide interpretable mental health screening via psychometric practices  \cite{lai2023psy, ohse2024gpt, radwan2024predictive, ravenda2025llms, rosenman2024llm}. These findings suggest that AI has the potential to bridge the gap between subjective and objective indicators, enabling more personalized and adaptive assessment, while also raising critical questions related to privacy, bias, and cross-cultural applicability.

\section{Methodology}

Our research employed a mixed-methods approach consisting of two main phases:

\subsection{Prototype Development Phase}
We developed a functional prototype of a natural language-based mental health assessment tool to provide a concrete reference point for expert evaluation. The prototype integrated the Symptom Checklist-90 (SCL-90) scale \cite{derogatis1973scl} into a conversational format using the Doubao 1.5 Pro large language model~\cite{doubao2023}on the Coze platform \cite{coze2023}. This phase aimed to transform abstract concepts into a tangible system that experts could evaluate.

\subsection{Expert Evaluation Phase}



We conducted semi-structured interviews with 11 mental health professionals, comprising six counselors and five psychiatrists. These interviews served multiple purposes: to gain insight into their current workflows and assessment methodologies, to explore their perspectives regarding the challenges associated with self-reported data, to gather feedback on our prototype system, and to identify essential design requirements for future development. Through these conversations with practitioners, we aimed to establish a comprehensive understanding of the clinical context in which our system would operate and the specific needs it should address.

\subsection{Data Analysis}

Interview transcripts were analyzed using thematic analysis to identify key patterns and insights. We focused on three main areas: (1) common mental health issues among college students, (2) professional inquiry approaches and diagnostic criteria, and (3) specific feedback on our prototype system.

\section{Prototype Design and Development}

\subsection{Introduction to the System}



To address our research questions, we developed a prototype system using advanced language models, providing interview participants with a concrete demonstration of natural language interaction for mental health assessment. Our system transforms traditional psychological testing into a conversational experience by integrating standardized assessments within natural dialogue.

The prototype is built on the Doubao 1.5 Pro large language model deployed on the Coze platform, selected for its strong performance in Chinese language understanding and contextual conversation management. We implemented the SCL-90 as the underlying assessment instrument for two reasons: it is widely used for mental health screening among college students, and its 90 items can be effectively restructured into conversational content, allowing the system to elicit symptom information through natural dialogue rather than explicit questioning.

The system adaptively navigates through assessment areas based on user responses, focusing on domains where potential concerns are detected while maintaining a supportive conversational flow.

\subsection{System Architecture and Workflow}

The system architecture consists of four core modules that form a complete assessment loop. The natural language interface converts standardised SCL-90 questions into three types of natural inquiry patterns: situational questions corresponding to interpersonal sensitivity dimensions, experiential questions corresponding to somatic symptom dimensions, and coping questions assessing psychological resilience. The dynamic assessment engine is capable of analysing emotional intensity and symptom keywords in user language in real time, constructing a symptom matrix, and intelligently selecting follow-up questions based on current symptom coverage to evaluate psychological resilience.The context processing mechanism maintains conversational coherence through a three-layer memory architecture to ensure a natural and smooth assessment process. The feedback and intervention module generates three-dimensional feedback based on the results of the assessment and provides the relevant support resources when risk signals are detected.

\begin{figure}[htbp]
   \centering
     \includegraphics[width=3.4in]{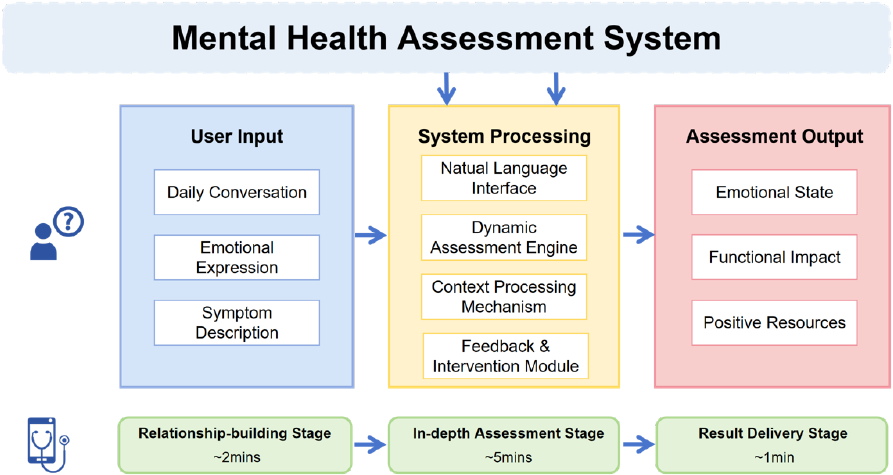}
     \vspace{-1.0em}
     \caption{System Workflow}
     \label{fig:System Workflow}
\end{figure}

The system workflow is divided into three stages (Figure \ref{fig:System Workflow}). The first stage, which takes approximately two minutes, is the relationship-building stage. This stage establishes trust through everyday topics and perform basic determination of the degree of impairment in social functioning . The in-depth assessment phase (approximately five minutes) dynamically activates relevant symptom dimension assessment pathways based on the user's initial feedback. The results delivery stage (approximately one minute) analyses user feedback to generate multi-dimensional assessment results, including emotional state, functional impact and positive resources.

\subsection{Large Language Model Configuration}

On Coze platform, we made several key configuration changes to the Doubao 1.5 Pro model. These included setting randomness parameter to 0.8 to ensure dialogue variety and naturalness while maintaining response consistency; employing a three-round dialogue memory mechanism to balance dialogue coherence and computational complexity; and setting maximum response length to 4, 096 tokens to enable the system to generate detailed evaluation content.

We constructed a multilevel instruction structure to clarify the system's professional identity as a 'college student mental health counsellor', used the first person to conduct naturalistic dialogues for the assessment, and implemented a dynamic assessment system, naturalistic questioning techniques, and a hybrid feedback mechanism. This mechanism describes the three assessment phases and their implementation strategies in detail, incorporating the nine dimensions of the symptoms and their scoring criteria of the SCL-90. These nine dimensions and their scoring criteria are integrated into the dialogue logic to provide the system with a professional assessment basis.

We designed a language conversion mechanism that transforms standardized scale items into natural dialogue expressions, aiming to preserve the essence of the assessment while improving the naturalness of communication. For example, the standard question \textit{'Do you often feel nervous or easily agitated?'} is converted into \textit{'How do you react when you are under stress lately?'}. This approach reduces the time required for assessment to 5-10 minutes compared to traditional standardized psychological assessments, while potentially increasing user engagement and data authenticity. All procedures follow established ethical codes for psychological assessment.

\vspace{-1mm}

\section{User Interview Progress}

In order to achieve the research objectives, we recruited university counsellors and psychiatrists, with whom we conducted semi-structured in-depth interviews.

Table ~\ref{tab:participants} categorises participants by profession and displays their basic and professional information, including gender, specialty field, occupation (Psychological Counselor = Psy, Psychiatrist = Ali), years of experience (years), clinical hours dedicated to mental health cases (hours), and department affiliation (Department). The use of a hyphen (-) indicates that the information is not applicable or was not provided by the participant.


\begin{table}[htbp]
  \caption{Participant Information}
  \label{tab:participants}
  \resizebox{0.5\textwidth}{!}{
  \begin{threeparttable}
    \begin{tabular}{|c|c|c|c|c|c|c|}
      \hline
      \textbf{ID} & \textbf{Sex} & \textbf{Professional field} & \textbf{Occ.} & \textbf{Yrs} & \textbf{Hours} & \textbf{Dept.} \\
      \hline
      P1 & F & College students' mental health & Psy & 5 & 500 & - \\
      \hline
      P2 & F & College students' mental health & Psy & 10 & 300+ & - \\
      \hline
      P3 & F & College students' MH, emotional disorders, stress & Psy & 6 & 1500 & - \\
      \hline
      P4 & M & College students' MH and stress management & Psy & 15 & 3000 & - \\
      \hline
      P5 & F & College students' MH, emotional disorders, stress & Psy & 3 & 1000 & - \\
      \hline
      P6 & F & MH, emotional disorders, stress management & Psy & 12 & 5000 & - \\
      \hline
      S1 & F & Other (mental disorders) & Ali & 10 & - & Psy \\
      \hline
      S2 & M & Dysthymic disorder & Ali & 10 & - & Psy \\
      \hline
      S3 & F & MH, emotional disorders, stress management & Ali & 13 & - & PC \\
      \hline
      S4 & F & College students' mental health & Ali & 16 & - & Psy \\
      \hline
      S5 & F & College students' MH, emotional disorders, stress & Ali & 6 & - & Psy \\
      \hline
    \end{tabular}
  \end{threeparttable}
  }
\end{table}

\subsection{Recruitment of Subjects}

Participants were recruited through an online questionnaire. They were recruited from tertiary hospitals and screened according to the years of experience and treatment approach. Finally, 11 participants (six counsellors and five psychiatrists) were recruited to conduct online interviews after signing the informed consent form. After the interview, all participants received a remuneration of 100 yuan.

\subsection{Interview Question Design}

In order to understand the processes and principles of psychological counselling and diagnosis for college students, and to further enhance the adaptive mental health assessment system, two open-ended interview questionnaires have been designed targeting participants' professions, based on the functionalities of the developed prototype and the goals for improvement. 

The questionnaires progressed in stages, starting with a basic introduction to understand college students' psychological status and collecting objective information. We then collected information on counselling and diagnostic processes and entry points. Together with the participants, we refined the model and rationale. Finally, we collected feedback and information on professionalisation and improvement. The questionnaire encouraged a comprehensive approach to students' psych problems, to achieve an accurate classification, understand assessment indicators, and collect effective counselling and diagnostic strategies.

\subsection{Interview Process}

The interviews were scheduled according to each participant's preferred time and online platform, each lasting between 25 and 45 minutes. Participants were then informed about the recording process before the interviews officially began. The interviews focused on the mental health status of college students and were based on the questionnaire questions. The key key questions and related topics were then explored. Finally, researchers provided participants with relevant requirements for the AI usage, offering additional references and targeted feedback for participants.After the interviews were transcribed, the researchers corrected them and summarised the information thoroughly.

\subsubsection{Common Types of Mental Disorders Among College Students and Information Required for Diagnosis}

Psychiatrists and counsellors experience mild to moderate anxiety and depression among college students. They agree that understanding a client's(\textit{A person who seeks professional psychological services (e.g., therapy, assessment), particularly in nonmedical settings like counselling or psychotherapy. This term emphasizes collaboration and autonomy.} ) complaint and symptoms (severity, duration, etc.) and their circumstances (academic performance, interpersonal relationships, etc.) guides the conversation and observes the client's behaviour. This information is then supplemented by scales to assess the diagnosis.

The difference lies in the fact that counsellors' assessments are primarily based on psychological assessment criteria, while psychiatrists' judgments are formed based on diagnostic systems. For example, P2 states: \textit{"After thoroughly understanding the impact of the issues the client is facing, such as duration, severity, and the extent of impact on social functioning, we combine their subjective experiences to assess whether the issue falls under general psychological problems or severe psychological problems."}; S2 states, \textit{"We assess whether clients meet standardized diagnostic criteria (such as ICD-10), integrating hospital systems to construct core symptom clusters based on initial presentations. These clusters are progressively refined and differentiated to prioritize symptom onset, ultimately forming a diagnosis when meeting criteria for course and severity. Diagnostic priority is given to the distress level of symptoms."}

\subsubsection{Inquiry Approach and Assessment Diagnostic Criteria}

Psychological counselors typically adopt an inquiry approach centered on open communication. After clarifying the primary complaint, they primarily listen and provide emotional support, conducting superficial emotional cognition, behavioral assessment, mental state assessment, and crisis assessment through language. After jointly confirming the counseling direction with the client, they gradually implement intervention measures based on the information obtained. Their assessment criteria primarily include the SAS, SDS, SCL-90, and DSM-V.

\begin{quote}
\textit{"By analyzing potential issues or scenarios, I help students organize or explore what they've said. I may share my perspective with them and invite them to discuss whether we can delve deeper into that aspect together."}\textup{(P5)}
\end{quote}

Psychiatrists, after understanding the chief complaint, focus on confirming basic information and related symptoms. They construct a symptom cluster based on the symptoms described by the client, determine the focus, and then conduct relevant inquiries. Finally, they use open-ended questions to further understand the symptoms and determine the treatment plan. Their diagnostic criteria primarily include ICD-10, state-related knowledge, disease-related knowledge, and DSM-5.

\begin{quote}
\textit{"During the first consultation, we inquire about relevant circumstances, such as where they are currently attending school, their current environment, their childhood state, and their parents' parenting style. Then, we discuss issues which cause them confusion. We typically ask them to provide a specific scenario, and then we delve into the details of the event, using this as a starting point."}\textup{(S5)}
\end{quote}

It is noteworthy that during actual interviews, researchers once asked participants about their “questioning process.” However, most counselors and psychiatrists indicated that there is no fixed questioning process. When interacting with clients, they follow the principle of “following the client's train of thought,” adjusting subsequent questions based on the client's responses and their own professional orientation. They aim to guide clients toward further directed interpretations rather than posing isolated, unrelated, or closed-ended questions solely for information gathering. This response emphasises the distinct nature of addressing psychological symptoms, thereby significantly enhancing the research methodology. This is the rationale behind the selection of this title.

\subsubsection{Feedback on Prototype}

After standardized AI usage, counselors and doctors provided targeted feedback. All participants affirmed the AI's basic functions but also identified areas requiring improvement or enhancement.

Psychological counselors emphasized empathy and guidance, highlighting issues with the AI's empathy functionality and the coherence of its responses. The AI demonstrated preliminary empathy during use, establishing a foundation for comfort, but it still faced challenges such as rigid response language and insufficient recognition capabilities, which caused frustration for users.

For example, P1 state\textit{"During communication, it showed concern for my role as an anxious student, and the opening remarks were also natural." }Similarly, P3 noted that the AI's responses helped users understand their own state and provided relevant professional knowledge, but also mentioned issues with its ability to connect questions: \textit{"During a single-round interaction, it showed concern for my well-being, which made me feel at ease. However, when I entered simple, conversational responses, it immediately displayed “Unrecognized” and offered technical jargon, demanding a detailed description of the situation to complete the query. Yet, if I told it I didn't want to answer, it would switch to another topic and continue the conversation—which proved quite useful."}

Psychiatrists emphasize accuracy and comprehensiveness, highlighting AI's ability to capture and collect information. The AI adheres to psychological expertise and scale-based methodologies to ensure comprehensive information gathering. However, it exhibits issues with inaccurate responses and weak logical reasoning during data input, resulting in reduced consistency during usage.

S1 mentioned the issue of questioning approach: \textit{"It's a good listening partner. However, when I mentioned that I've been feeling down lately and have some troubles, it kept asking me about the disease-related aspects and didn't address my emotions."}

\vspace{-1mm}

\section{Discussion}

Our research aimed to develop and evaluate an LLM-based mental health assessment tool that overcomes limitations of traditional questionnaires through imperceptible assessment in natural conversation. This section discusses our findings, limitations, and future directions in relation to our core contribution: creating an intelligent agent that can conduct psychological assessments through natural dialogue while maintaining clinical validity.

\subsection{Summary of Contributions}

The primary contribution of this work is the development and initial validation of an intelligent mental health assessment system based on the Doubao 1.5 Pro large language model. Unlike traditional questionnaire-based assessments that often feel clinical and impersonal, our system transforms standardized SCL-90 assessment items into natural conversational elements. For example, rather than directly asking "Do you often feel nervous or easily agitated?",our system might inquire, "How do you react when you are under stress lately?" This approach maintains assessment validity while significantly improving the user experience.

Our semi-structured interviews with 11 participants (six counsellors and five psychiatrists) validated this approach, with participants noting that natural language interaction could potentially increase user engagement and data authenticity compared to traditional assessment methods.However, the interviews also revealed three areas in need of improvement before such systems can be used in clinical settings: the ability to show empathy, the accuracy of assessments, and crisis intervention protocols.

\subsection{Areas for Improvement}

\textbf{Empathy.} Research indicates that individuals with psychological symptoms or disorders often require communication from an equal stance with appropriate emotional support rather than excessive sympathy \cite{Wambsganss2022}. This finding was corroborated by the results of the clinician interviews conducted for this study. Some participants emphasised that users should "feel that their inner thoughts are understood and be given the courage to overcome negative emotions" \textup{(S2)}when interacting with the assessment system.

The current prototype demonstrates limitations in recognizing and appropriately responding to emotional cues, particularly when users express distress indirectly. To address this limitation, we propose enhancing our system with improved emotional detection capabilities and response templates derived from therapeutic communication practices \cite{Baumer2012}. Specifically, the system should initiate conversations from everyday topics, prioritize emotional comfort, and transition to supportive statements when detecting sustained emotional distress, such as \textit{"I notice this seems difficult to talk about. Perhaps we could take a moment to focus on something more comfortable before returning to this topic if you're willing."}

\textbf{Intelligent assessment.} The accuracy of psychological assessment depends significantly on the interactivity of language and comprehensiveness of information collection \cite{Rose2008}. Our current prototype faces challenges in recognizing certain response patterns and maintaining conversational coherence—limitations identified by multiple clinicians in our interviews.

To improve assessment accuracy, we will enhance the system's natural language understanding capabilities to better interpret colloquial expressions of psychological distress. For instance, when a user states, "I just don't feel like getting out of bed anymore," the system should recognize this as potentially indicating depression rather than physical fatigue. Additionally, we will implement more sophisticated conversational strategies to encourage elaboration without appearing intrusive, such as using open-ended prompts like \textit{"Would you like to share how that experience affected you?"} rather than direct assessment questions. These improvements directly address clinicians' concerns about the system's ability to gather clinically relevant information through natural conversation.

\textbf{Crisis intervention.} Effective crisis intervention capabilities are essential for any mental health assessment tool, as highlighted by both existing literature \cite{Pendse2020} and our interview participants. Current prototype lacks mechanisms for identifying and responding to indicators of acute psychological distress or suicidal ideation.

Based on clinician feedback, we propose implementing a specialized crisis detection module that can recognize linguistic patterns associated with imminent risk. When concerning language is detected, the system will implement a tiered response protocol: first providing supportive statements, then offering crisis resources, and finally activating an alert to the user's pre-designated support person in severe cases (with prior consent). This approach balances immediate safety concerns with user autonomy and privacy—a balance that multiple clinicians in our study identified as crucial for technology-mediated mental health assessment.

\subsection{Limitations and Future Work}

Despite the evidence from our research that demonstrates the potential of mental health assessments based on large language models, there are several key limitations that must be addressed.

Firstly, it is important to note that conversational rephrasing has the potential to alter the psychometric properties of the original assessment tool. This, in turn, may necessitate the undertaking of comparative validation studies in the future. Secondly, the system's adaptive approach to user responses challenges the integrity of the assessment, necessitating exploration of methods to balance natural conversation with psychometric integrity.

Presently, the prototype is undergoing expert interview testing, with more rigorous validation to be conducted during the forthcoming clinical trial evaluation phase. Concurrently, we will proceed with the exploration of multimodal assessment methods with a view to enhancing accessibility for diverse populations.

In summary, the findings of the present study demonstrate that mental health assessment using natural conversation based on large language models is technically feasible and has potential value in improving user engagement and data authenticity. However, before such systems can be deployed responsibly in clinical settings, there is a need for significant improvements in empathy, assessment accuracy, and crisis intervention capabilities. These findings provide a solid foundation for the further development of mental health assessment methods that can enhance the assessment experience while maintaining clinical efficacy. 

\section{Conclusion}

This study demonstrates the feasibility and potential benefits of a mental health assessment system based on natural language interaction. The integration of AI-driven technology with psychological assessment principles has been demonstrated to facilitate a more conversational and adaptive assessment process. A thematic analysis of semi-structured interviews with 11 participants (six counsellors and five psychiatrists) was conducted, leading to the identification of three fundamental design principles. The findings of this study indicate that AI-driven adaptive questioning provides a more comfortable and flexible assessment experience while demonstrating good effectiveness in improving psychological logs and assisting in the construction of psychological models. Given the individual willingness to use AI systems, despite privacy concerns, the information obtained through this method appears highly credible, highlighting its theoretical potential for practical psychological assessment applications. While further validation is required, this work may result in the development of novel assessment methods and the optimisation of counselling and diagnostic processes between mental health professionals and their clients.

\begin{acks}
This work is supported by the Natural Science Foundation of China (Grant No. 62302252).
\end{acks}

\end{document}